# NiO$_x$/β-Ga$_2$O$_3$ Heterojunction Diode Achieving Breakdown Voltage >3 kV with Plasma Etch Field-Termination

Yizheng Liu[1,a], Saurav Roy[1], Carl Peterson[1], Arkka Bhattacharyya[1], and Sriram Krishnamoorthy[1,a]

[1]Department of Materials, University of California Santa Barbara, Santa Barbara CA 93106, USA

a) Author to whom correspondence should be addressed. Electronic mail: yizhengliu@ucsb.edu and sriramkrishnamoorthy@ucsb.edu

*Abstract*: This work reports the fabrication and characterization of a NiO$_x$/β-Ga$_2$O$_3$ heterojunction diode (HJD) that uses a metallic nickel (Ni) target to deposit NiO$_x$ layers via reactive RF magnetron sputtering and lift-off processing with >3 kV breakdown voltage, record-low reverse current leakage under high reverse bias, and high junction electric fields (>3.34 MV/cm). The heterojunction diodes are fabricated via bilayer NiO$_x$ sputtering followed by self-aligned mesa-etching for field-termination on both large (1-mm$^2$) and small area (100-μm diameter) devices. The HJD exhibits a ~135 A/cm$^2$ forward current density at 5 V with a rectifying ratio of ~10$^{10}$. The minimum differential specific on-resistance is measured to be 17.26 mΩ•cm$^2$. The breakdown voltage on 100-μm diameter pads was measured to be greater than 3 kV with a noise floor-level reverse leakage current density (10$^{-8}$ ~10$^{-6}$ A/cm$^2$) until 3 kV, accomplishing a parallel-plane junction electric field to be at least 3.34 MV/cm at 3 kV with a power figure of merit (PFOM) >0.52 GW/cm$^2$. Temperature-dependent forward current density-voltage (J-V) measurements are performed from room temperature (25 ºC) to 200 ºC which showed a temperature coefficient of resistance (α) equaling 1.56, higher than that of β-Ga$_2$O$_3$ Schottky barrier diodes (SBDs), indicating potential conductivity degradation within NiO$_x$ at elevated temperatures.

Power devices are essential building blocks for efficient energy conversion applications in power electronic systems and are advancing rapidly driven by the development of wide bandgap (WBG) and ultra-wide band gap (UWBG) semiconductors and their devices[1,2]. Beta-gallium oxide (β-Ga$_2$O$_3$) is a promising power semiconductor because of its high critical electric field (E$_C$~8 MV/cm) and shallow hydrogenic n-type dopants[3–5]. By taking considerations of both incomplete ionization effect and material purity level (compensation), β-Ga$_2$O$_3$ currently exhibits the highest power figure of merit (PFOM) among many other UWBG semiconductors as gallium nitride (GaN), silicon carbide (SiC), and aluminum nitride (AlN)[6]. However, due to the difficulty of achieving reliable p-type conductivity in β-Ga$_2$O$_3$, high breakdown voltage devices with E$_C$ approaching β-Ga$_2$O$_3$'s theoretical limit are often realized in forms of heterojunction structures with proper edge termination (ET). To substitute p-type β-Ga$_2$O$_3$, nickel oxide (NiO$_x$) is commonly adopted as an alternative to fabricate heterojunction vertical power devices[7–12] on n-type β-Ga$_2$O$_3$ drift layer grown via halide vapor phase epitaxy (HVPE) and metalorganic chemical vapor deposition (MOCVD)[13]. Rapid emergence of NiO$_x$/β-Ga$_2$O$_3$ heterojunction diodes (HJD) suggests that such combination is possible to reach high breakdown voltage (V$_{br}$), and high parallel-plane junction electric field with potential avalanche capability[14].

Across current literatures[7–12], NiO$_x$ is deposited on β-Ga$_2$O$_3$ via RF magnetron sputtering of compound NiO target. Unlike conventional extrinsically doped compound semiconductors, stoichiometric nickel oxide (NiO) is a Mott-Hubbard insulator[15]. To impart conductivity to NiO, it is suggested to flow excess amount of oxygen gas (O$_2$) under high vacuum condition in combination with argon gas (Ar) during sputtering to deposit $NiO_x$ that is stoichiometrically imbalanced between Ni and $O^{15}$. The excess $O_2$ introduces divalent nickel ion vacancy ($V_{Ni}''$) via oxidation of $Ni^{2+}$ to $Ni^{3+}$ to invoke vacancy-mediated transport[16].

This work demonstrates the fabrication of $NiO_x/\beta\text{-}Ga_2O_3$ ET heterojunction diodes with $NiO_x$ deposited via RF reactive magnetron sputtering of metallic Ni target. The large area device (1-mm$^2$) exhibits a catastrophic breakdown voltage of ~1.5 kV, and on-state current of 1.12 A. Small area device (100-μm diameter) breakdown voltage exceeds 3 kV, with a punch-through parallel plane junction critical electric field greater than 3.34 MV/cm, minimum differential specific on-state resistance ($R_{on,sp}$) of 17.26 mΩ•cm$^2$, and PFOM >0.52 GW/cm$^2$. Reverse leakage of the small area device remains ultra-low ($10^{-8}$ ~$10^{-6}$ A/cm$^2$) at 3 kV.

The main fabrication steps for the $NiO_x/\beta\text{-}Ga_2O_3$ vertical HJD are illustrated in **Fig. 1(a)** with the schematic view of the fabricated device shown in **Fig. 1(b)**. The process starts with a standard solvent clean (acetone, isopropanol alcohol, and de-ionized (DI) water for 3 minutes under sonication in each solvent) followed by a 5-minute 49.98% hydrofluoric (HF) acid treatment, and DI water rinse. A photoresist lift-off mask is then patterned using optical lithography to prepare the sample for bilayer $NiO_x$ RF magnetron reactive sputtering. During the sputter deposition, a ~30 nm relatively resistive ($R_{sheet}$~215 kΩ/□) $NiO_x$ layer is first deposited by using a 99.999% purity Ni metallic target under oxygen-deficient conditions (Ar/$O_2$: 22/10 sccm) with a 4.5 mTorr chamber pressure at 150 W RF power and a deposition rate of 0.375 nm/min. Next, a ~30 nm relatively conductive ($R_{sheet}$~7-9 kΩ/□) $NiO_x$ layer is continuously grown under oxygen-rich condition (Ar/$O_2$: 10/8 sccm) with a 4 mTorr chamber pressure at the same RF power (0.476 nm/min). Sheet resistance of each $NiO_x$ layer is verified on a sapphire monitor wafer via Hall measurement. After the $NiO_x$ deposition, a ~30 nm of Ni metallic Ohmic anode cap is sputtered on top of the $NiO_x$ stack *in-situ* under pure Ar condition (Ar:25 scccm) without breaking the vacuum at 150 W and 5 mTorr to minimize $NiO_x$ surface contamination under ambient exposure. Thickness of $NiO_x$ layers under different sputter conditions are confirmed and corroborated via profilometer and X-ray reflectivity (XRR) measurements. After sputtered Ni capping, sample is transferred to e-beam evaporation immediately for metallization of Au/Ni (30/50 nm) stack. The 50-nm Ni serves as a hard mask for subsequent self-aligned $\beta\text{-}Ga_2O_3$ mesa etching. The bi-layer $NiO_x$[7] and metal cap are then lifted off in a heated NMP solution. The now exposed $\beta\text{-}Ga_2O_3$ drift layer is then etched ~ 1 μm below the heterojunction interface using a self-aligned inductively coupled $BCl_3$ (20 sccm) plasma dry etch at 200 W to enable rounded/curved corners[17], shown as the scanning electron microscopy (SEM) images in shown in **Figs. 1(c)** and **1(d)**, to prevent electric field crowding at the heterojunction interface. Finally, a backside Ti/Au (80/200 nm) Ohmic metal stack is deposited via e-beam evaporation to conclude the device fabrication.

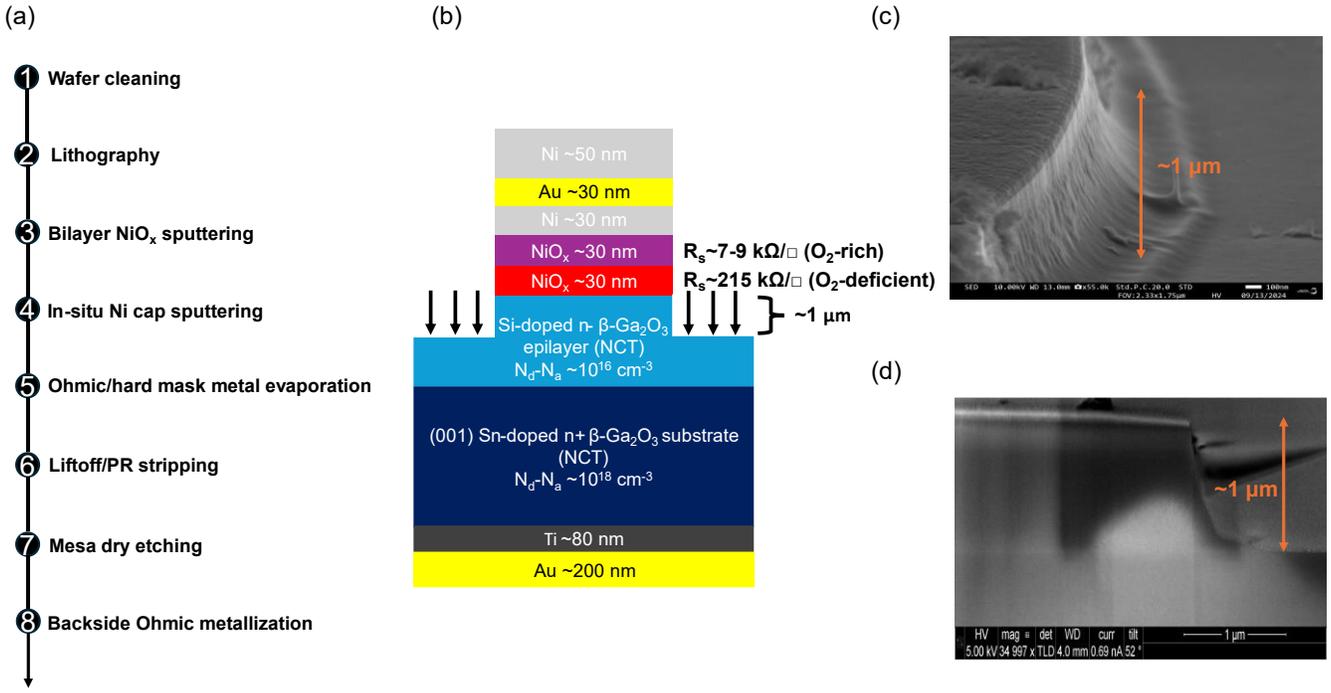

FIG. 1. (a) Main fabrication steps for the mesa-etched NiO$_x$/β-Ga$_2$O$_3$ vertical HJD. (b) Schematic view of mesa-etched bilayer NiO$_x$/β-Ga$_2$O$_3$ vertical HJD. (c) Scanning electron microscopy (SEM) image of the circular device's edge after plasma etching. (d). Cross-sectional SEM image of the circular device's edge with curved corner after plasma etching.

The wafer consists of a ~650 μm thick β-Ga$_2$O$_3$ (001) substrate (Sn doping: ~10$^{18}$ cm$^{-3}$) along with a ~10 μm HVPE Si-doped drift layer from Novel Crystal Technology (NCT). The average net doping and thickness of the HVPE drift region are extracted to be ~9.39×10$^{15}$ cm$^{-3}$ and ~11.9 μm from high-voltage capacitance-voltage (C-V) measurements on 1-mm$^2$ pad at 1 MHz, as shown in **Figs. 2(a)** and **2(b)** using a relative permittivity of 12.4 of β-Ga$_2$O$_3$[18]. The device reaches punch-through at a reverse voltage of >1.1 kV during the high-voltage C-V measurement when absolute capacitance of the HJD flattens out and no longer changes as a function of applied reverse bias.

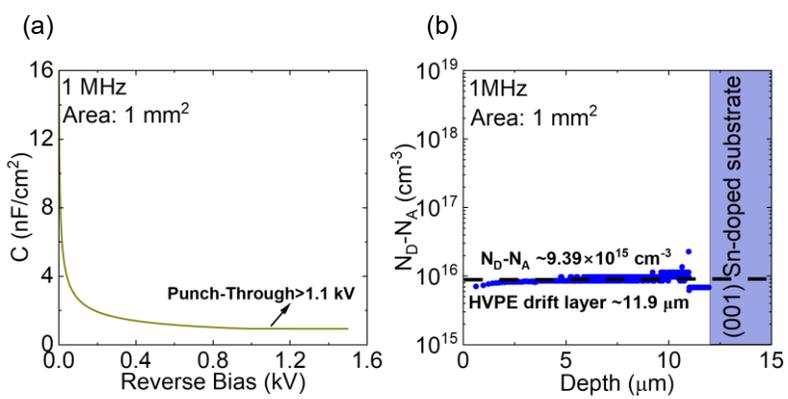

FIG. 2. (a) High-voltage capacitance-voltage (C-V) characteristics on 1-mm$^2$ pad at 1MHz. (b) Extracted apparent charge density vs. Depth profile of the mesa-etched NiO$_x$/β-Ga$_2$O$_3$ vertical HJD.

The forward linear J-V characteristics of the 100-μm diameter of mesa-etched HJDs are shown in **Fig 3(a)** with the minimum $R_{on,sp}$ is extracted to be 17.26 mΩ·cm$^2$ at ~4 V as shown in the inset. The device turns on at ~2.6 V with an on-state current density of 135 A/cm$^2$ at 5 V. The reverse leakage of the HJD is at ~10$^{-8}$ A/cm$^2$ under low reverse bias (-4 V), exhibiting a rectifying ratio of ~10$^{10}$, shown in **Fig 3(b)**. Current density is estimated by including a 45° angle current spreading in the 10.9-μm unetched drift region[19].

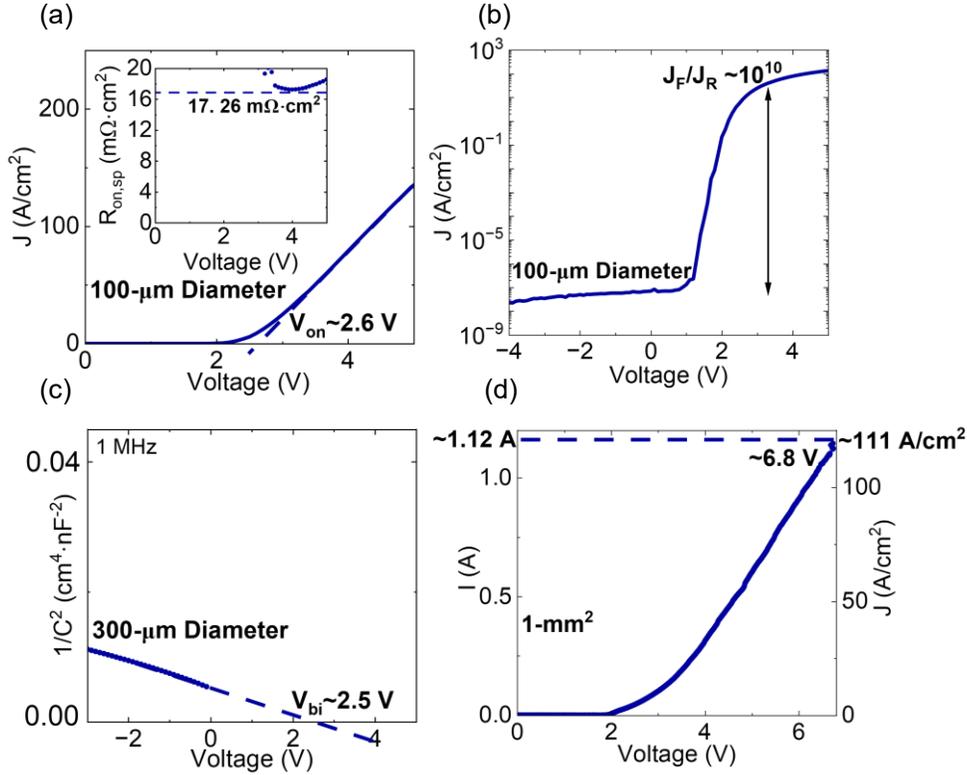

FIG. 3. (a) Linear forward J-V characteristics of 100-μm diameter heterojunction diodes. Inset is the differential specific ON resistance ($R_{on,sp}$) of the heterojunction diode. (b) J-V characteristics in log scale. (c) $1/C^2$ vs. Voltage at 1 MHz (300-μm diameter). (d) I-V and J-V characteristics of the large-area heterojunction diode (1-mm$^2$).

The built-in potential ($V_{bi}$) of the NiO$_x$/β-Ga$_2$O$_3$ junction is extracted to be ~2.5 V from $1/C^2$ vs. Voltage characteristics (**Fig 3(c)**) at 1 MHz, consistent with the turn-on voltage extracted from J-V characteristics. Additionally, large area device (1-mm$^2$) exhibited a maximum forward current of 1.12 A at 6.8 V with corresponding current density of ~111 A/cm$^2$ from pulsed DC current-voltage (I-V) measurement with a pulse width of 50-μs and a pulse period of 0.5 ms (10% Duty Cycle) in **Fig 3(d)**.

The reverse leakage and breakdown characteristics of eight NiO$_x$/β-Ga$_2$O$_3$ mesa-etched HJDs (rinsed in Silicone oil) on small area pads (100-μm diameter) are shown in **Fig 4(a)**. The reverse leakage current remains below noise level of the measurement tool limit (10$^{-8}$~10$^{-6}$ A/cm$^2$, nA) under high reverse bias, and all devices show no sign of breakdown up to 3 kV at tool's reverse bias limit.

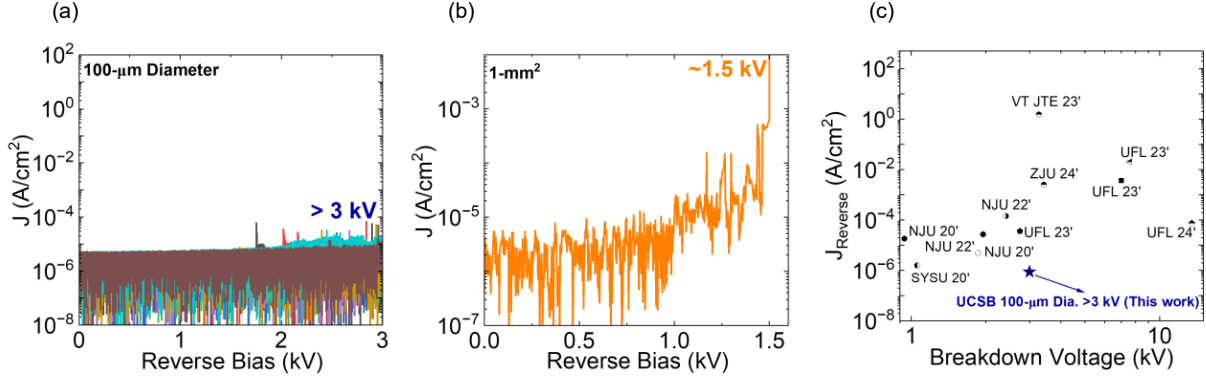

FIG. 4. (a) Breakdown characteristics of the 100-μm diameter heterojunction diodes. (b) Breakdown characteristics of the 1-mm² heterojunction diode. (c) Benchmarking of reverse leakage ($J_{Reverse}$) vs. breakdown voltage of $NiO_x/\beta$-$Ga_2O_3$ heterojunction diode[7–12,20–23].

The parallel-plane junction electric field ($E_C$) of the HJDs on 100-μm is extracted using Eqn. (1)[24].

$$BV_{PT} = E_C W_{D,PT} - \frac{qN_{D,PT}W_{D,PT}^2}{2\varepsilon_s} \qquad (1)$$

For breakdown voltages of ($BV_{PT}$) >3 kV, punch-through depletion width ($W_{D,PT}$) of ~11.9 μm, drift layer average doping ~9.39×10¹⁵ cm⁻³, and relative vacuum permittivity ($\varepsilon_s$) of β-$Ga_2O_3$ at 12.4$\varepsilon_0$ for (001) orientated β-$Ga_2O_3$[18], we extract the parallel-plane junction electric field ($E_C$) to be >3.34 MV/cm. It should be noted that we are not making any assumptions regarding doping and thickness of the HVPE layer, but accurately extracting this information from high voltage C-V measurements to precisely estimate the parallel-plane junction electric field at the $NiO_x/\beta$-$Ga_2O_3$ junction.

Moreover, the breakdown voltage of the large area (1-mm²) HJD (rinsed in Silicone oil) is ~1.5 kV with the lowest noise-floor level reverse leakage current density of ~10⁻⁶ A/cm² before device experiences catastrophic breakdown, as shown in **Fig 4(b)**. The $NiO_x/\beta$-$Ga_2O_3$ 100-μm diameter mesa-etched heterojunction diode, fabricated via RF reactive sputtering from pure metallic Ni target, exhibits the lowest reverse leakage current density at its breakdown voltage (> 3 kV) compared to literature reported $NiO_x/\beta$-$Ga_2O_3$ HJDs, according to the benchmarking of reverse leakage vs. breakdown voltage[7–12,20–23] shown in **Fig 4(c)**, which is possibly attributed to the HF rinse cleaning before device fabrication, and the *in-situ* Ohmic anode capping in the sputter chamber, preserving the interfacial quality of HJDs.

Moving to larger 300-μm diameter devices, **Fig 5** illustrates the temperature dependent J-V characteristics for mesa-etched $NiO_x/\beta$-$Ga_2O_3$ heterojunction diodes. The extracted turn-on voltage of devices decreases from 2.1 V to 1.8 V as temperature increases from room temperature (300 K) to 200 °C (473 K). The current density of devices also decreases as measurements are carried out at elevated temperatures. The inset of **Fig 5** describes the dependence of differential on-resistance ($R_{on}$) of HJDs and absolute temperature. By fitting $R_{on}$ using power law expression $R_{on} = R_{on}^{300\,K}(T/300)^\alpha$ shown in **Fig 5** inset, the temperature coefficient of on-resistance (α) of the mesa-etched $NiO_x/\beta$-$Ga_2O_3$ HJD is extracted to be 1.56, which is higher than α values (0.87,0.73)[17,25] reported in literatures for β-$Ga_2O_3$ Schottky barrier diodes (SBDs), suggesting possible conductivity degradation with increasing temperature in sputtered $NiO_x$[26].

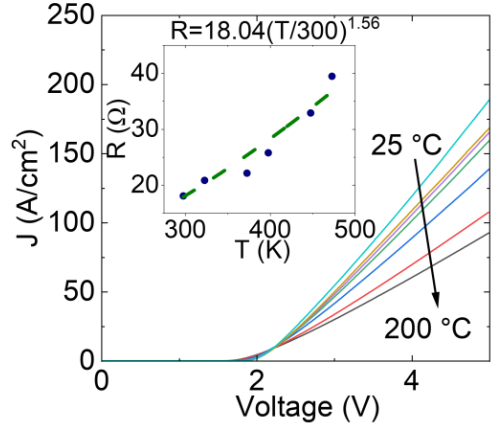

FIG. 5. Temperature dependent J-V characteristics of the 300-μm diameter heterojunction diode. Inset shows ON-resistance as a function of temperature.

The small area HJD (100-μm diameter) with a $R_{on,sp}$ of 17.26 mΩ•cm$^2$ and a breakdown voltage exceeding 3 kV accomplishes a PFOM of at least 0.52 GW/cm$^2$, demonstrating the feasibility of NiO$_x$/β-Ga$_2$O$_3$ HJDs fabricated from reactive sputtering via pure metallic Ni target. Improvements in the PFOM can be made via reduction of differential specific on-resistance by using thinner n-type β-Ga$_2$O$_3$ bulk substrate underneath the n-type HVPE drift region.

In summary, we demonstrated NiO$_x$/β-Ga$_2$O$_3$ based power device by introducing bilayer NiO$_x$ reactive sputtering via metallic Ni target, *in-situ* Ohmic anode capping, and plasma etch field-termination. The small area devices (100-μm diameter circular pads) exhibit very low reverse leakage current density of 10$^{-8}$~10$^{-6}$ A/cm$^2$ up until 3 kV, accompanied by a $R_{on,sp}$ of 17.26 mΩ•cm$^2$, accomplishing a PFOM greater than 0.52 GW/cm$^2$. The parallel-plane junction electric field of the HJD is accurately extracted to be > 3.34 MV/cm via high-voltage C-V measurements without making assumptions on apparent doping densities or HVPE drift region thickness. By using Ni metallic target, the entire device fabrication process is carried out under room temperature condition without involving any high-temperature processing techniques to accomplish > 3 kV NiO$_x$/β-Ga$_2$O$_3$ diode breakdown performance. The large area (1-mm$^2$) device showcases an absolute forward current of 1.12 A at ~6.8 V. Additionally, the extracted temperature coefficient of resistance (α = 1.56) for NiO$_x$/β-Ga$_2$O$_3$ mesa-etched heterojunction diode, higher than β-Ga$_2$O$_3$ SBDs (α = 0.73), indicating potential conductivity degradation within sputtered NiO$_x$.

## ACKNOWLEDGMENTS

The authors acknowledge funding from the ARPA-E ULTRAFAST program (DE-AR0001824) and Coherent/II-VI Foundation Block Gift Program. A portion of this work was performed at the UCSB Nanofabrication Facility, an open access laboratory. Y. Liu would like to acknowledge the SEM imaging assistance from K. Chanchaiworawit.

## DATA AVAILABILITY

The data that support the findings of this study are available from the corresponding authors upon reasonable request.

## REFERENCES

[1] A.J. Green, J. Speck, G. Xing, P. Moens, F. Allerstam, K. Gumaelius, T. Neyer, A. Arias-Purdue, V. Mehrotra, A. Kuramata, K. Sasaki, S. Watanabe, K. Koshi, J. Blevins, O. Bierwagen, S. Krishnamoorthy, K. Leedy, A.R. Arehart, A.T. Neal, S. Mou, S.A. Ringel, A. Kumar, A. Sharma, K. Ghosh, U. Singisetti, W. Li, K. Chabak, K. Liddy, A. Islam, S. Rajan, S. Graham, S. Choi, Z. Cheng, and M. Higashiwaki, "β-Gallium oxide power electronics," APL Materials **10**(2), 029201 (2022).

[2] M. Buffolo, D. Favero, A. Marcuzzi, C. De Santi, G. Meneghesso, E. Zanoni, and M. Meneghini, "Review and outlook on GaN and SiC power devices: industrial state-of-the-art, applications, and perspectives," IEEE Transactions on Electron Devices, (2024).

[3] A.J. Green, K.D. Chabak, E.R. Heller, R.C. Fitch, M. Baldini, A. Fiedler, K. Irmscher, G. Wagner, Z. Galazka, and S.E. Tetlak, "3.8-MV/cm Breakdown Strength of MOVPE-Grown Sn-Doped β-$Ga_2O_3$ MOSFETs," IEEE Electron Device Letters **37**(7), 902–905 (2016).

[4] A. Kuramata, K. Koshi, S. Watanabe, Y. Yamaoka, T. Masui, and S. Yamakoshi, "High-quality β-$Ga_2O_3$ single crystals grown by edge-defined film-fed growth," Japanese Journal of Applied Physics **55**(12), 1202A2 (2016).

[5] A.T. Neal, S. Mou, S. Rafique, H. Zhao, E. Ahmadi, J.S. Speck, K.T. Stevens, J.D. Blevins, D.B. Thomson, and N. Moser, "Donors and deep acceptors in β-Ga2O3," Applied Physics Letters **113**(6), (2018).

[6] Y. Zhang, and J.S. Speck, "Importance of shallow hydrogenic dopants and material purity of ultra-wide bandgap semiconductors for vertical power electron devices," Semiconductor Science and Technology **35**(12), 125018 (2020).

[7] H.H. Gong, X.H. Chen, Y. Xu, F.-F. Ren, S.L. Gu, and J.D. Ye, "A 1.86-kV double-layered NiO/β-$Ga_2O_3$ vertical p–n heterojunction diode," Applied Physics Letters **117**(2), (2020).

[8] X. Lu, X. Zhou, H. Jiang, K.W. Ng, Z. Chen, Y. Pei, K.M. Lau, and G. Wang, "1-kV Sputtered p-NiO/n-$Ga_2O_3$ Heterojunction Diodes With an Ultra-Low Leakage Current Below 1∼μA/$cm^2$," IEEE Electron Device Letters **41**(3), 449–452 (2020).

[9] M. Xiao, B. Wang, J. Spencer, Y. Qin, M. Porter, Y. Ma, Y. Wang, K. Sasaki, M. Tadjer, and Y. Zhang, "NiO junction termination extension for high-voltage (> 3 kV) $Ga_2O_3$ devices," Applied Physics Letters **122**(18), (2023).

[10] J.-S. Li, C.-C. Chiang, X. Xia, H.-H. Wan, F. Ren, and S.J. Pearton, "7.5 kV, 6.2 GW $cm^{-2}$ NiO/β-$Ga_2O_3$ vertical rectifiers with on–off ratio greater than $10^{13}$," Journal of Vacuum Science & Technology A **41**(3), (2023).

[11] X. Xia, J.-S. Li, C.-C. Chiang, F. Ren, and S.J. Pearton, "Fabrication and Device Performance of 2.7 kV/2.5 A NiO/Ga$_2$O$_3$ Heterojunction Power Rectifiers," ECS Transactions **111**(2), 103 (2023).

[12] J. Wan, H. Wang, C. Zhang, Y. Li, C. Wang, H. Cheng, J. Li, N. Ren, Q. Guo, and K. Sheng, "3.3 kV-class NiO/β-Ga$_2$O$_3$ heterojunction diode and its off-state leakage mechanism," Applied Physics Letters **124**(24), (2024).

[13] A. Bhattacharyya, C. Peterson, K. Chanchaiworawit, S. Roy, Y. Liu, S. Rebollo, and S. Krishnamoorthy, "Over 6 μm thick MOCVD-grown low-background carrier density (10$^{15}$ cm$^{-3}$) high-mobility (010) β-Ga$_2$O$_3$ drift layers," Applied Physics Letters **124**(1), (2024).

[14] F. Zhou, H. Gong, M. Xiao, Y. Ma, Z. Wang, X. Yu, L. Li, L. Fu, H.H. Tan, and Y. Yang, "An avalanche-and-surge robust ultrawide-bandgap heterojunction for power electronics," Nature Communications **14**(1), 4459 (2023).

[15] Y. Hong, X. Zheng, H. Zhang, Y. He, T. Zhu, K. Liu, A. Li, X. Ma, W. Zhang, J. Zhang, and Y. Hao, "Oxygen Stoichiometry Engineering in P-Type NiO$_x$ for High-Performance NiO/Ga$_2$O$_3$ Heterostructure p–n Diode," Physica Rapid Research Ltrs, 2400109 (2024).

[16] S. Nandy, U.N. Maiti, C.K. Ghosh, and K.K. Chattopadhyay, "Enhanced p-type conductivity and band gap narrowing in heavily Al doped NiO thin films deposited by RF magnetron sputtering," Journal of Physics: Condensed Matter **21**(11), 115804 (2009).

[17] S. Roy, B. Kostroun, J. Cooke, Y. Liu, A. Bhattacharyya, C. Peterson, B. Sensale-Rodriguez, and S. Krishnamoorthy, "Ultra-low reverse leakage in large area kilo-volt class β-Ga$_2$O$_3$ trench Schottky barrier diode with high-k dielectric RESURF," Applied Physics Letters **123**(24), (2023).

[18] A. Fiedler, R. Schewski, Z. Galazka, and K. Irmscher, "Static dielectric constant of β-Ga$_2$O$_3$ perpendicular to the principal planes (100),(010), and (001)," ECS Journal of Solid State Science and Technology **8**(7), Q3083 (2019).

[19] S. Roy, A. Bhattacharyya, C. Peterson, and S. Krishnamoorthy, "2.1 kV (001)-β-Ga$_2$O$_3$ vertical Schottky barrier diode with high-k oxide field plate," Applied Physics Letters **122**(15), (2023).

[20] J.-S. Li, H.-H. Wan, C.-C. Chiang, T.J. Yoo, M.-H. Yu, F. Ren, H. Kim, Y.-T. Liao, and S.J. Pearton, "Breakdown up to 13.5 kV in NiO/β-Ga$_2$O$_3$ vertical heterojunction rectifiers," ECS Journal of Solid State Science and Technology **13**(3), 035003 (2024).

[21] Y. Wang, H. Gong, Y. Lv, X. Fu, S. Dun, T. Han, H. Liu, X. Zhou, S. Liang, and J. Ye, "2.41 kV Vertical P-NiO/n-Ga$_2$O$_3$ Heterojunction Diodes With a Record Baliga's Figure-of-Merit of 5.18 GW/cm$^2$," IEEE Transactions on Power Electronics **37**(4), 3743–3746 (2021).

[22] F. Zhou, H. Gong, W. Xu, X. Yu, Y. Xu, Y. Yang, F. Ren, S. Gu, Y. Zheng, and R. Zhang, "1.95-kV beveled-mesa NiO/β-Ga$_2$O$_3$ heterojunction diode with 98.5% conversion efficiency and over million-times overvoltage ruggedness," IEEE Transactions on Power Electronics **37**(2), 1223–1227 (2021).

[23] J.-S. Li, H.-H. Wan, C.-C. Chiang, T.J. Yoo, F. Ren, H. Kim, and S.J. Pearton, "NiO/Ga$_2$O$_3$ Vertical Rectifiers of 7 kV and 1 mm$^2$ with 5.5 A Forward Conduction Current," Crystals **13**(12), 1624 (2023).

[24] B.J. Baliga, *Fundamentals of Power Semiconductor Devices* (Springer Science & Business Media, 2010).

[25] M. Xiao, B. Wang, J. Liu, R. Zhang, Z. Zhang, C. Ding, S. Lu, K. Sasaki, G.-Q. Lu, and C. Buttay, "Packaged Ga$_2$O$_3$ Schottky rectifiers with over 60-A surge current capability," IEEE Transactions on Power Electronics **36**(8), 8565–8569 (2021).

[26] J.-S. Li, H.-H. Wan, C.-C. Chiang, F. Ren, and S.J. Pearton, "Annealing Stability of NiO/Ga$_2$O$_3$ Vertical Heterojunction Rectifiers," Crystals **13**(8), 1174 (2023).